\documentclass[conference]{IEEEtran}

\usepackage{listings}
\usepackage{algorithm}
\usepackage{algorithmic}
\usepackage{times}
\usepackage{mathtools}
\usepackage{moreverb}
\usepackage{pst-node}
\usepackage{graphicx}
\usepackage{multirow}
\usepackage{amssymb}
\usepackage{subfig}
\usepackage{balance}

\lstdefinestyle{MyListStyle}{
language=C, % choose the language of the code 
basicstyle=\ttfamily\footnotesize, % the size of the fonts used  
tabsize=3, % sets default tabsize to 2 spaces 
captionpos=b, % sets the caption-position to bottom 
breaklines=true, % sets automatic line breaking 
breakatwhitespace=false, % sets if automatic breaks should only happen at whitespace 
escapeinside={\%*}{*)}, % if you want to add a comment within your code 
frame=single, % adds a frame around the code 
numbers=left,
numbersep=5pt,
numberstyle=\tiny,
}
\newtheorem{theorem}{Theorem}
\newtheorem{lemma}{Lemma}
\newtheorem{definition}{Definition}
\newtheorem{claim}{Claim}

\newcommand{\CVI}{\ensuremath{\mathit{CVI(\Upsilon)}}}
\newcommand{\CVIonly}{\ensuremath{\mathit{CVI}}}

\usepackage{amsmath}
\usepackage{xspace}
\newcommand{\ENTRY}{\text{\em ENTRY}\xspace}
\newcommand{\EXIT}{\text{\em EXIT}\xspace}
\begin{document}

\title{Sliced Slices: Separating Data and Control Influences}

%\numberofauthors{2} %  in this sample file, there are a *total*
% of EIGHT authors. SIX appear on the 'first-page' (for formatting
% reasons) and the remaining two appear in the \additionalauthors section.
%

\author{
\IEEEauthorblockN{Shrawan Kumar}
\IEEEauthorblockA{Tata Research Design \\
    and Development Centre\\
%Tata Consultancy Services Ltd, India\\
Tata Consultancy Services Ltd,\\
Pune, India\\
Email: shrawan.kumar@tcs.com}\\
\and
\IEEEauthorblockN{Amitabha Sanyal}
\IEEEauthorblockA{Department of Computer Science \\
    and Engineering \\
%Indian Institite of Technology Bombay, India\\
Indian Institute of Technology Bombay, \\
India\\
       Email: as@cse.iitb.ac.in}\\
\and
\IEEEauthorblockN{Uday P. Khedker}
\IEEEauthorblockA{Department of Computer Science \\
    and Engineering \\
%Indian Institite of Technology Bombay, India\\
Indian Institute of Technology Bombay, \\
India\\
       Email: uday@cse.iitb.ac.in}
}
%\author{
%\alignauthor
%Shrawan Kumar\\
%       \affaddr{Tata Research Design and Development Centre\\
%       Tata Consultancy Services Ltd, India}\\
%       \email{shrawan.kumar@tcs.com}

% 2nd. author
%\alignauthor
%Amitabha Sanyal\\
%       \affaddr{Department of Computer Science and Engineering \\
%           Indian Institite of Technology Bombay, India}\\
%       \email{as@cse.iitb.ac.in}
%Uday P. Khedker\\
%       \affaddr{Department of Computer Science and Engineering \\
%           Indian Institite of Technology Bombay, India}\\
%       \email{uday@cse.iitb.ac.in}
%}

\maketitle

\maketitle
%\tableofcontents

\begin{abstract}
Backward slicing has been used extensively in program understanding, debugging
and scaling up of program analysis. 
For large programs, the size of the conventional backward slice 
is about 25\% of the program size. This may be too large to be 
useful. 
Our investigations reveal that in
general, the size of a slice is influenced more by computations
governing the control flow reaching the slicing criterion than by the
computations governing the values relevant to the slicing criterion.
We distinguish between the two by defining data slices and control 
slices both of which are smaller than the conventional slices which
can be obtained by combining the two.
This is useful because for many applications, the individual data or control 
slices are sufficient.

Our experiments show that for more than 50\% of cases, the data
slice is smaller than 10\% of the program in size. 
%{\red (10\% of the conventional slice or 10\% of the program?)} 
Besides, the time to compute data or control slice is comparable to 
that for computing the conventional slice.
\end{abstract}

 %0

%A category including the fourth, optional field follows...
%\category{F.3.2}{Semantics of Programming Languages}[Program Analysis]
% A category with the (minimum) three required fields
%\category{D.3.4}{Processors}{Compilers}

%\terms{Algorithms, Languages, Performance}

%\keywords{Slicing, data slice, control slice, program understanding}

\section{Introduction}
Program slicing, introduced in 1984 by 
Mark Weiser~\cite{PROGRAM_SLICING}, 
strips down a large program to a smaller version 
based on the requirements of program observation.
Many variants of slices such as 
forward, backward, dynamic,
and abstract slice etc. have been devised~\cite{SLICING_SURVEY_SILVA}. They have
been used for different purposes 
like program understanding, debugging, testing, maintenance, software
quality assurance and reverse engineering among others. 
A brief description of various applications of program slices is given 
by Binkley et. al.~\cite{SLICING_SURVEY_BINKLEY}. 

\begin{figure*}[!htb]
\begin{minipage}{0.30\linewidth}
\subfloat[Backward slice]{%
\begin{lstlisting}[style=MyListStyle]
int main()
{
  int i=0,j=0, k=0, st;
  int l, t=0, u, v;
  l=get_input();
  while (fn1(l)) 
    t+=2;
  while (i<1000) 
  {
    i= i+ fn2(); 
    st = fn3();
    if (st ==1)
      { j++; k++; }
    else { j+=2; k+=1; }
    u = j-k;
    if (t>100) 
      v = u ; 
  }
  return 0;
}
int fn1(); // a complex function 
int fn2(); // a complex function 
int fn3(); // some function 
\end{lstlisting}%
}
\end{minipage}
\hspace{0.5cm}
\begin{minipage}{0.30\linewidth}
\subfloat[Data slice]{%
\begin{lstlisting}[style=MyListStyle]
int main()
{
  int j=0, k=0, st;
  int u, v;
  while (*) 
  {
    st = fn3();
    if (st ==1)
      { j++; k++; }
    else { j+=2; k+=1; }
    u = j-k;
    if (*) 
      v = u ;
  }
  return 0;
}
int fn3(); // some function 
\end{lstlisting}%
}
\end{minipage}
\hspace{0.5cm}
\begin{minipage}{0.30\linewidth}
\subfloat[Control slice]{%
\begin{lstlisting}[style=MyListStyle]
int main()
{
  int i=0;
  int l, t=0, u, v;
  l=get_input();
  while (fn1(l)) 
    t+=2;
  while (i<1000) 
  {
    i= i+ fn2(); 
    if (t>100) 
      v = u ; 
  }
  return 0;
}
int fn1(); // a complex function 
int fn2(); // a complex function 
\end{lstlisting}%
}
\end{minipage}
\caption{Usual backward slice, data slice and control slice}
\label{fig:datacrt-example}
\end{figure*}

Among static slicing techniques, backward slicing 
which answers the question ``which program statements can influence the
given variables at the given statement?'' seems more natural and is most common.
It identifies the portion
of program that one would be interested in while understanding a
computation or debugging for an erroneous output. In safety property
checking, rather than verifying the property on whole program, one
can verify the property on the static backward slice with respect to
slicing criterion derived from the specific property of interest. As a
result, backward slice helps scaling up of property checking techniques
also. 
%Consequently, static slice in general, and backward static slice
%in particular, is the most widely used slicing technique in the area
%of program comprehension, testing, debugging, maintenance, program
%analysis, program differencing etc.

While static slicing is efficient and scalable, the size of the computed
slice may remain a matter of concern. Empirical studies~\cite{SLICE_SIZES} have shown that size of a static slice on an
average is about 30\% of the program size. For large programs, this size
could be too large. Our investigations of the factors influencing the size of a
slices reveal that most statements are
included in a slice due to some conditions governing the reachability
of the statement involved in the slicing criteria rather than due to
the values of variables in the slicing criteria. 
These statements are irrelevant for understanding
the computations leading to the values of the variables. 

%\subsection{Motivating example}
As a motivating example, assume
that the program in Figure~\ref{fig:datacrt-example}(a) computes
an erroneous value of \texttt{u} at line 17.
It is obvious that for slicing criterion $\langle 17,
\mbox{\texttt{u}}\rangle $, no static slicing can reduce the program anymore. Therefore
using the conventional backward slice is of no help in getting a reduced program for debugging this
program.
A careful examination reveals that the value of \texttt{u} does not depend on
the values of variables \texttt{i} or \texttt{t}. These variables are used in the
conditions which decide the reachability of line 17 in the execution.
Since we know that a wrong value is getting computed at line 17,
reachability of line 17 is obvious and need not be established. Thus,
computations of \texttt{i} and \texttt{t} are irrelevant to our purpose.

Figure~\ref{fig:datacrt-example}(b) shows a portion of the program which
is sufficient to understand the computation of \texttt{u} and to debug the
reason for its wrong value. Any erroneous statement responsible for
an erroneous value has to be contained in this program fragment. The
statements that have been removed only alter the reachability of line 17
and not the value computed for \texttt{u}. Note that functions \texttt{fn1} and \texttt{fn2}
have been sliced out as they are not required any longer.

Observe that retaining the program structure requires that a
conditionally executed statement in the original program must also be
included as a conditionally executed statement in the program slice.
However, since the values of variables appearing in the condition are
irrelevant, we replace conditions by `*' which stands for a random
value chosen from $\{true, false\}$ when it is executed.

It is easy to see that the resulting slice is much smaller in comparison
to conventional backward slice (incidentally the whole program in this
case) and still sufficient to debug (or understand) the computation of
\texttt{u} at the slicing criterion. We call such a slice as {\em data slice}.

Consider a contrasting requirement of debugging the program when line 17
is not getting executed a desired number of times. For this purpose we
only need to see how the reachability of line 17 is getting influenced.
How the value of \texttt{u} is computed is irrelevant. Therefore we need to
know how values of \texttt{i} and \texttt{t} are getting computed as they appear in
the conditions that govern whether or not line 17 will be reached. We
show the portion of code in Figure~\ref{fig:datacrt-example}(c). It is
sufficient to understand when line 17 in original program (mapped to
line 12 in slice) is reachable. Functions \texttt{fn1} and \texttt{fn2} have to be
part of this program portion while function \texttt{fn3} is not required. We
call such a slice as {\em control slice}.

In addition to debugging and program understanding, this separation
of concerns is helpful in property checking during program verification because
it reduces the size of the program to be examined.
First we derive a slicing criterion $SC$ from the property to be checked, 
compute the corresponding data slice and check the
property on the slice. If the property holds then we are done as
the property will hold in original program also. If the property does
not hold, we compute a control slice with respect to program point
related to property in question and check the reachability of this point
in the control slice. If it is not reachable then property
holds in original program.\footnote{However, if it is found
reachable then the answer is not straight forward.}

The contributions of this paper are:
%We make the following contributions: 
We define the concepts of data  and control slices, relate them to the
conventional slices, provide an  algorithm to compute data and control
slices and show its  soundness. We also provide empirical measurements
on real life  programs that show that the data  slices are indeed much
smaller than complete slices and are computable in comparable time.
 %1
\section{Background}
\subsection{Variants of Program Slices}

The classical backward slicing tries to find the program fragment
that influences a slicing criterion.  Forward
slicing~\cite{FORWARD_SLICING} discovers the statements that are
influenced by a given slicing criterion. Chopping~\cite{CHOPPING}
discovers statements influenced by a source criterion on paths to a
target criterion. A dynamic slice~\cite{DYNAMIC_SLICING} computes a
subset of program statements which affect a slicing criterion in a
particular run of the program. Assertion 
slicing~\cite{ASSERTION_SLICING,PRECOND_SLICING,WP_SLICING} 
is a technique which computes set of
statements which are sufficient to ensure a post condition or 
the statements which will be executed starting from a given pre
condition. All these variants use control and data influences in an
integrated manner. To the best of our knowledge there is no work which
distinguishes between data and control influences.

\subsection{Control flow graph and data dependence}

\textbf{Program model}. We present our ideas in context of 
imperative programs modeled in terms of assignment statements, 
conditional statements, {\em while} loops, 
and procedure calls. We also allow {\em break} and {\em continue} 
statements in loops. Without any loss of generality,
we restrict ourselves to goto-less 
programs with {\em single-entry loops} and two-way branching
conditional statements at the source level.  
%%To keep the
%%discussion simple, we do not consider more general conditional statements
%%like {\em switch-case} or other loop constructs like {\em for, repeat-until,
%%do-while}. But this does not take away generality of our ideas as 
%%the excluded constructs can be realised by the included ones.

\textbf{Control Flow Graph (CFG)}. 
We use the standard notion of 
{\em control flow graph (CFG)} $G = \langle N, E\rangle $ where $N$ is the set of nodes, $E$ 
is a set of directed edges in $N \times N$~\cite{DATAFLOW_FRAMEWORK}.
$\ENTRY$ and $\EXIT$ are distinguished 
nodes representing the entry and exit of program. We use $s \rightarrowtail t$ 
and $s \stackrel {l} {\rightarrowtail} t$ to denote unconditional and 
conditional edges respectively, where 
$l\in \{true, false\}$ indicates the branch outcome. 
We assume that two special edges $\ENTRY \rightarrowtail \EXIT$ and $\EXIT \rightarrowtail \EXIT$ are added in CFG.
There is a {\em one-to-one} correspondence between nodes of CFG and statements of the program, hence
we will use the terms statement and node interchangeably.

\textbf{Data dependence}. 
A definition $d$ of a variable $v$ in node $p$
is said to be {\em reaching definition}~\cite{DATAFLOW_FRAMEWORK} for a node point $q$, if 
there is a control flow path from $p$ to $q$ devoid of any other definition of $v$.
A variable $x$
at location $l$ is said to be {\em data dependent} on a definition  
$d$ of $x$, if $d$ is {\em reaching definition} for $l$. 
The set of definitions of variables in $X$ reaching $l$ is denoted by
$DU(l,X) = \{ d \mid \exists v \in X. v$ at $l$ is data dependent on $d\}$.
We will use $REF(t)$ to denote set of variables whose value is referred in a statement $t$.

\subsection{Program states and traces}
Let $V$ be set of all variables in program $P$ 
and $\Re$ be
the set of all values which the variables can take in $P$.
\begin{definition}{(Program state)}.
A program state is  a valuation of all variables in the program at a given 
instant during program execution. It is represented by a map $\theta$
: $V \rightarrow \Re$ such that 
$\theta(v)$ denotes the value of 
$v \in V$ in the program state $\theta$. 
\end{definition}

\begin{definition}{(Restricted program state)}.
Given $X \subseteq V$, 
a {\em X-restriction} of program state $\theta$,  
  denoted as $\lfloor \theta \rfloor_X$, is a map $X \rightarrow \Re$ 
  such that $\forall x \in X . \lfloor \theta \rfloor_X (x) = \theta(x)$.
\end{definition}
%We define a state of program execution
%as follows:
\begin{definition}{(Execution state)}.
An execution state is a pair $\langle n, \theta\rangle $ where $\theta$ is a program state
and $n$ is a CFG node. 
\end{definition}

Execution of a program can be seen as a sequence of
execution states starting with execution state $\langle \ENTRY, \sigma_0\rangle $ 
where $\sigma_0$ is initial program state. 
The subsequent execution state $\langle n^{'}, \sigma^{'}\rangle $ for a given execution state
$\langle n, \sigma\rangle $ is decided by semantics of statement
corresponding to node $n$ and program state $\sigma$. Let function $TRAN(\langle n, \sigma \rangle)$ provide the subsequent execution state of $\langle n, \sigma \rangle$.

\begin{definition}{(Trace)}.
A (possibly infinite) sequence of execution states $[\langle n_i, \sigma_i\rangle],\; i\geq 0$ is said to be a 
 trace, provided %the following hold:
%\begin{itemize}
%\item $n_0=\ENTRY$ and $\sigma_0$ is given initial program state
$n_0=\ENTRY$ and $\sigma_0$ is given initial program state, and
%\item $\forall i \geq 0 : \langle n_{i+1}, \sigma_{i+1}\rangle = TRAN(\langle n_i, \sigma_i \rangle)$
$\forall i \geq 0 : \langle n_{i+1}, \sigma_{i+1}\rangle = TRAN(\langle n_i, \sigma_i \rangle)$.
%\end{itemize}
\end{definition}
When the trace sequence is finite and ends with an 
execution state $\langle \EXIT, \theta \rangle$ then the trace is 
called a terminating trace.
Unless stated otherwise, a trace means 
a terminating trace in the rest of this paper.
\subsection{Post-dominance and control dependence}
Backward slicing algorithms are implemented efficiently using
post-dominance and control dependence~\cite{PDG, INTERPROC_SLICING}. 
\begin{definition}{(Post-dominance)}.
A node $n_2$ {\em post-dominates} a node $n_1$ if every path 
from $n_1$ to $\EXIT$ contains $n_2$. If, in addition $n_1 \neq n_2$ then 
$n_2$ is said to {\em strictly post-dominate} $n_1$.
\end{definition}
\begin{definition}{(Control dependence)}.
A node $n_3$ is {\em control dependent} on an edge 
$n_1 \stackrel {l} {\rightarrowtail} n_2$ if
%\begin{itemize}
%\item $n_3$ {\em post-dominates} $n_2$, and
$n_3$ {\em post-dominates} $n_2$, and
%\item $n_3$ does not strictly {\em post-dominates} $n_1$.
$n_3$ does not strictly {\em post-dominate} $n_1$.
%\end{itemize}
\end{definition}
Later Podgurski and Clarke~\cite{WEAK_CD} 
introduced concept of
{\em strong post-dominance} and {\em weak control dependence},
to consider
execution of a statement being dependent on loop termination. The
previous definition of post-dominance and control dependence were termed 
as 
{\em weak post-dominance} and {\em strong control dependence} respectively.
\begin{definition}{(Strong post-dominance)}.
A node $n_2$ {\em strongly post-dominates} a node $n_1$ if every infinite 
path starting at $n_1$ contains $n_2$. If in addition, $n_1 \neq n_2$ then 
$n_2$ {\em strictly strongly post-dominates} $n_1$.
\end{definition}
\begin{definition}{(Weak control dependence)}.
A node $n_3$ is {\em weakly control dependent} on an edge 
$n_1 \stackrel {l} {\rightarrowtail} n_2$ if
%\begin{itemize}
%\item $n_3$ {\em strongly post-dominates} $n_2$, and
$n_3$ {\em strongly post-dominates} $n_2$, and
%\item $n_3$ does not strictly {\em strongly post-dominates} $n_1$.
$n_3$ does not strictly {\em strongly post-dominate} $n_1$.
%\end{itemize}
\end{definition}

%If $n_3$ is weakly (strongly) control dependent on an edge 
%$n_1 \stackrel{l}{\rightarrowtail} n_2$ then we also say that $n_3$ is 
%weakly (strongly) {\em l-control} dependent or just weakly (strongly) control dependent on the node 
%$n_1$.
%Weak control dependence is also transitive.

Only $\EXIT$ node strongly post-dominates a loop exit edge. 
Therefore, all the nodes that weakly post-dominate a loop condition
are weakly control dependent on 
loop exit edge.
Bilardi and Pingali~\cite{COMP_WEAK_CD} 
give efficient algorithms for computing strong post-dominance and weak 
control dependence relationship.
%We use {\em sconds(t)} to denote set of 
%conditions on which statement $t$ is {\em strongly control dependent}.
%{\color {magenta}
For our purpose, we will need to know whether a statement is controlled
by a condition through a chain of weak/strong control dependence. For this,
we define a transitive closure of weak/strong control dependence.
%we define a closure of strong control dependence for statement $t$ is 
%denoted by {\em sconds*(t)}.
\begin{definition}{(Transitive control dependence)}.
For given statement $s$ and condition $c$, if there is a path $\pi$ 
in PDG from $c$ to $s$ consisting of only (weakly/strongly) control 
dependent edges then we say that $s$ is 
{\em transitive control dependent} on $c$, 
written as $c \rightsquigarrow s$.
If $\pi$ consists of only {\em strong control dependent} edges then we say
$s$ is {\em strongly transitive control dependent} on $c$, 
written as $c \longrightarrow s$. 

If $\pi$ starts with a control dependent edge labeled as $e$ then we qualify the transitive control dependence with edge $e$ as  
$c \stackrel {e} {\rightsquigarrow} s$ or  
$c \stackrel {e} {\longrightarrow} s$. Obviously,
$c \stackrel {e} {\longrightarrow} s \implies c \stackrel {e} 
{\rightsquigarrow} s$.
\end{definition}

Following properties are obvious, for the programs under our discourse.
\begin{enumerate}
\item[SP1]
$(c \stackrel {e_1} {\longrightarrow} s \wedge 
        c \stackrel {e_2} {\longrightarrow} s) \implies e_1 = e_2$
\item[SP2] $c \stackrel {e} {\longrightarrow} s \implies$ no path to $s$ from an edge
$\bar {e} \neq e$ of $c$ can bypass 
$e$. 
\end{enumerate}
%}

\subsection{Subprogram and backward slice}
An important requirement of a slice is that the behaviour of the slice
must be a specified subset of the original program's behaviour. A
subset of original program's behaviour is specified through a pair
$\Upsilon=\langle l, V\rangle $, known as {\em slicing criterion},
where $l$ is a statement location and $V$ is set of variables. It is
interpreted as values of variables $V$ just before executing statement
at $l$. We will use $\Upsilon$ and $\langle l, V\rangle$ 
interchangeably to denote a slicing criterion. Where context
is clear, we will use $l$ and $V$ to denote the location and variables
set components, respectively. We will use $LV(t)$ to denote 
the slicing criterion $\langle t, REF(t)\rangle$.

A \textit{subprogram} of a program $P$ is a program carved out of $P$ by 
deleting some statements such that program structure remains
intact in that for each statement $n$ that appears in {\em subprogram}, if $n$ is 
enclosed by a condition in $P$, then $n$ must be enclosed by a condition
in {\em subprogram} too. 
%{\red (Have removed notation $P^S$ from here.)} 

An \textit{augmented program} ($P^A$) is the result of transforming a 
given program $P$ for a slicing criterion $\Upsilon=\langle l,V\rangle$ by 
inserting a $SKIP$ statement at location $l$.
Obviously, an augmented program is equivalent to the original program. 
To compute a slice of program $P$ with respect to 
$\Upsilon$, we compute the slice $S$ of the augmented program with respect to $\Upsilon$ 
with the restriction that the inserted $SKIP$ statement is retained
in $S$. Now with $l$ standing for $SKIP$ statement, $\Upsilon$ can be seen
as specification for the desired subset of original program's behaviour for $P$, $P^{A}$ and $S$. 
Henceforth, we assume that $SKIP$ statement, inserted at location $l$, is part of every slice with respect to $\Upsilon$.
%{\red (Have removed forward reference to observation window.)} 

%{\color {magenta}
Henceforth, we will assume that each statement in $P$ is labeled uniquely
and the nodes in CFG are labeled with corresponding statement label. 
Statements in subprogram will get their label from the ones given in $P$.
As a result, in CFG, $G^s=\langle N^s, E^s \rangle$ constructed for a sub 
program $S$ of program $P$, having CFG, $G:\langle N, E \rangle$, 
$N^s \subseteq N$. To be more explicit, given a node $n^s \in N^s$ and 
$n \in N$, $n^s = n$ would mean that they represent statement with same 
label. This also holds for special nodes $\ENTRY$ and $\EXIT$.
%}

\begin{definition}{(SC execution state)}.
Given a program $P$ and slicing criterion $\langle l,V \rangle$, 
execution states of $P$ having statement 
location as $l$ are called {\em SC execution states}.
\end{definition}

%{\color {magenta}
Given a program $P$ and slicing criterion $\Upsilon=\langle l, V\rangle$, let 
$\tau:[(n_i, \sigma_i)], \; 0 \leq i \leq k$. 
be the trace for input $I$ with $m$ {\em SC-execution} states.
Let $\tau^s:[(n^s_i, \sigma^s_i)], \; 0 \leq i \leq k^s$, 
be trace for a slice $S$ of $P$, on same input $I$ with $m^s$ 
{\em SC-execution} states.
Let 
$[\langle n_{l_j}, \sigma_{l_j}\rangle], \; 1 \leq j \leq m$ 
and
$[\langle n^s_{l_j}, \sigma^s_{l_j}\rangle], \; 1 \leq j \leq m^s$ 
be the sequence of {\em SC-execution states} in 
in $\tau$ and $\tau^s$ respectively. 
We say,
$\langle n_{l_j}, \sigma_{l_j}\rangle$
and
$\langle n^s_{l_j}, \sigma^s_{l_j}\rangle$ for $0 < j < min(m, m^s)$
are {\em corresponding SC-execution} states.
Now the window of observation for $\Upsilon=\langle l, V\rangle$ is the sequence of {\em V-restricted} program states
$[\lfloor \sigma_{l_j} \rfloor_V], \; 1 \leq j \leq m$. 
We will call such a sequence as the  trace window of 
observation $TW(P,I,\Upsilon)$.
%}

\begin{definition}{(SC-equivalent trace)}.
A trace $\tau$ of $P$ on an input I is called {\em SC-equivalent} to trace
$\tau^s$
on slice $S$ for same input I, if $TW(P,I,\Upsilon) = TW(S, I, \Upsilon)$.
\end{definition}

We express the definition of a backward slice $P^B_{\Upsilon}$ for a program $P$ and a slicing 
criterion $\Upsilon$ as follows.

\begin{enumerate}
\item $P^B_{\Upsilon}$ is {\em subprogram} of $P^A$.
\item For every input I on which original program terminates, \\
    $TW(P, I, \Upsilon) = TW(P^B_{\Upsilon}, I, \Upsilon)$. 
\end{enumerate}

Of all methods that use some form of data flow to compute the backward
slice,  the  methods of  Ferrante  et  al.~\cite{PDG}  and Horwitz  et
al.~\cite{INTERPROC_SLICING} using  program dependence graphs (PDG)
produce the minimal slice.
 %2

 \section{Data and control  slices} A backward slice is  the answer to
the  question  ``Given  a  slicing  criterion $\Upsilon  =  \langle  l,
V\rangle$,  which program  statements  can affect  the  values of  the
variables  $V$?" This  question  can be  meaningfully  split into  two
parts:
\begin{enumerate}
\item[$Q1$]  Which  statements decide whether
program control will reach $l$?
\item[$Q2$]  Assuming that control reaches $l$, which
program statements decide the values of the variables
$V$.
\end{enumerate}

As      mentioned      in      the     motivating      example      of
Figure\ref{fig:datacrt-example},  often we  are  interested  in
the separate answers to  questions $Q1$  or $Q2$, even if we want to
use them together. We  call the
subprogram resulting from the answer  to $Q1$ as a {\em control slice}
and that to $Q2$ as a {\em data slice}.

\begin{figure*}[!htb]
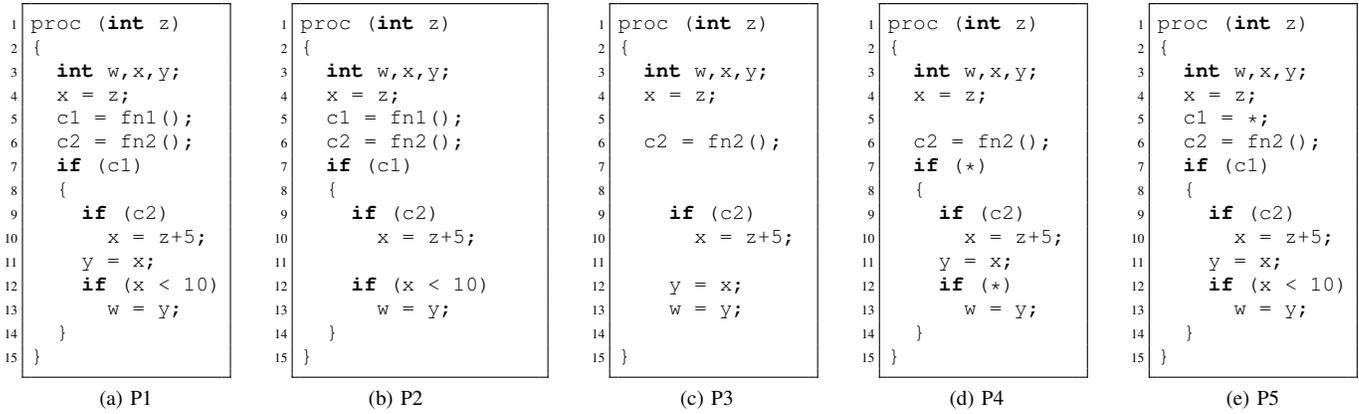

\begin{minipage}{1.00in}
\subfloat[P1]{%
\begin{lstlisting}[style=MyListStyle]
proc (int z)
{
  int w,x,y;
  x = z;
  c1 = fn1();
  c2 = fn2();
  if (c1)
  {
    if (c2)
      x = z+5;
    y = x;
    if (x < 10)
      w = y;
  }
}
\end{lstlisting}%
}
\end{minipage}
\hspace{0.8cm}
\begin{minipage}{1.25in}
\subfloat[P2]{%
\begin{lstlisting}[style=MyListStyle]
proc (int z)
{
  int w,x,y;
  x = z;
  c1 = fn1();
  c2 = fn2();
  if (c1)
  {
    if (c2)
      x = z+5;
      
    if (x < 10)
      w = y;
  }
}
\end{lstlisting}%
}
\end{minipage}
\hspace{0.8cm}
\begin{minipage}{1.00in}
\subfloat[P3]{%
\begin{lstlisting}[style=MyListStyle]
proc (int z)
{
  int w,x,y;
  x = z;
            
  c2 = fn2();
     

    if (c2)
      x = z+5;
                   
    y = x;
    w = y;
     
}
\end{lstlisting}%
}
\end{minipage}
\hspace{0.8cm}
\begin{minipage}{1.00in}
\subfloat[P4]{%
\begin{lstlisting}[style=MyListStyle]
proc (int z)
{
  int w,x,y;
  x = z;

  c2 = fn2();
  if (*)
  {
    if (c2)
      x = z+5;
    y = x;
    if (*)
      w = y;
  }
}
\end{lstlisting}%
}
\end{minipage}
\hspace{0.8cm}
\begin{minipage}{1.00in}
\subfloat[P5]{%
\begin{lstlisting}[style=MyListStyle]
proc (int z)
{
  int w,x,y;
  x = z;
  c1 = *;
  c2 = fn2();
  if (c1)
  {
    if (c2)
      x = z+5;
    y = x;
    if (x < 10)
      w = y;
  }
}
\end{lstlisting}%
}
\end{minipage}
\caption{Various forms of data slices}
\label{fig:ds-example}
\end{figure*}

In Figure~\ref{fig:ds-example},  P1 is  the original program  in which
the  codes for  the functions  \texttt{fn1} and  \texttt{fn2}  are not
shown.  P2 is the control slice for the slicing criterion $\langle 13,
\lbrace\mbox{\texttt{y}}\rbrace \rangle$.   The conditions \texttt{c1}
and \texttt{x}~$<$~10 at lines 7 and 12 decide whether program control
reaches line 13, and therefore they are part of the control slice.
Further, the  value of the condition \texttt{x}~$<$~10  may be decided
by the  assignment at line  10, which, in  turn, is controlled  by the
condition at  line 9. Therefore  both line 12  and line 10 are  in the
control slice.  For similar reasons, lines 4, 5, and 6 also have to be
included.

To obtain  the data slice  for P1 for  the same slicing  criterion, we
reason as  follows.  The value  of \texttt{y} at  line 13 is  given by
assignment at  11 and the condition at  line 12 has no  impact on this
value. Therefore line 11 is included  in the data slice but line 12 is
not. The value of  x assigned to y at line 11  is computed by lines 10
and 4.  Of  these, the value that reaches line  11 during execution is
decided by condition  at line 9. Therefore lines 4, 9  and 10 are also
included in  the data slice. Further,  line 6 is also  included, as it
computes the  value to be used in  the condition at line  9.  No other
statement affects  the value of  \texttt{y} at line 13.  The resulting
program P3 is the data slice.

We now formalize the notions of control and data slice. 

\subsection{Control slice}
A  control slice  of  a program  $P$  wrt.  to  the slicing  criterion
$\Upsilon=\langle  l,  V\rangle$,  denoted  $P^C_{\Upsilon}$,  is  the
backward slice of  $P$ with respect to the  slicing criterion $\langle
l, \emptyset\rangle$.

Thus,  the  sliced program  $P^C$  contains  those  statements of  the
original  program which merely  caused the  program control  reach the
program point $l$.  We now show that the control slice is contained in
the backward slice.
\begin{lemma}
\label{cs_subset_bs_lem}
$P^C_{\Upsilon} \subseteq P^B_{\Upsilon}$
\end{lemma}
\begin{IEEEproof}
From the definition of control slice, we have 
 $P^{C}_{\Upsilon}$ = $P^{B}_{\langle l,\emptyset \rangle}$. 
Further,  backward slices have the 
property~\cite{PROGRAM_SLICING} 
$V_1 \subset V_2$ implies
 $P^B_{\langle l, V_1 \rangle} \subset P^B_{\langle l, V_2 \rangle}$ 
and thus  $P^B_{\langle l, \emptyset 
\rangle} \subseteq P^B_{\langle l, V \rangle}$.
Consequently $P^C_{\Upsilon} \subseteq P^B_{\Upsilon}$
\end{IEEEproof}

Since computing backward slice is a well studied problem, we can
compute the control slice by computing the  backward slice with
respect  to  the  slicing  criterion  $\langle  l,  \emptyset  \rangle$
following any of the existing approaches.

\subsection{Data slice}
While  P3 is  an  answer to  the  question $Q2$  posed for  the
criterion $\langle 13, \lbrace \mbox{\texttt{y}}\rbrace \rangle$ and can therefore  be regarded as a
data  slice,  it  is   not  suitable  for  applications  like  program
understanding and  debugging.  For  example, the information  that the
statement  at line  13 may  not  be executed  at all  is missing  from
P3. Thus apart from the statements that decide on the values of
variables at a  slicing criterion, we also need  to include statements
that explicate  the paths along  which the computation of  such values
takes place.

Therefore, we also include  conditions that impact the reachability of
$l$.  This  is shown in P4,  and the included conditions  are shown as
`*' indicating  a non-deterministic  branch.  We make  such conditions
non-deterministic  because their  values are  inconsequential  for our
purpose.   Further, if these  conditions were  made concrete,  then we
would  also  have to  include  additional  statements affecting  their
values,  increasing   the  size   of  the  slice.    We  call   such  a
non-deterministic conditional  as {\em abstract  conditional}.  During
execution,  such a  conditional  can randomly  evaluate  to $true$  or
$false$.

The form of the slice as shown in P4 is good for program understanding
and  debugging.    However  it  falls  short  if   used  for  property
verification. The reason is  that while the {\em abstract conditional}
helps in keeping the size of  the sliced program small, it elides path
conditions that are important for verification. As an example, suppose
we want  to check  the property $\mbox{\texttt{y}}  < 20$ at  line 13.
This  property  holds  in  original  program  but  does  not  hold  in
P4. However, if  we retain the included conditionals  in concrete form
and abstract  out instead those  assignment statements that  assign to
the variables  involved in  the conditional, we  get P5.  The property
$\mbox{\texttt{y}}  <  20$  holds  for  this program.  Note  that  the
assignments to  \texttt{x} cannot be  eliminated since they also  determine the
value of the variables of the slicing criterion.

An {\em abstract  assignment} of the form \mbox{\texttt{x}}~=~* assigns  to \mbox{\texttt{x}} a random
value from  some domain. If  \mbox{\texttt{x}} is an  integer, for example, then  \mbox{\texttt{x}} is
assigned     an     integer     value     between     $-2^{31}$     to
$2^{31}-1$.\footnote{Assuming a 4 byte 2s complement representation.}

Due to inclusion of abstract conditions and assignments, there will be
multiple  execution paths  on a  given  input. In  general, the  slice
produced  by abstract assignments  will have  less number  of possible
executions paths for  same input in comparison to  the one produced by
abstract conditionals.  For example, P4 may have 4 different execution
paths on a given input while P5  will have only two such paths on same
input.  As  a result, the slice produced through abstract assignments
is more useful  in property checking. The flip  side is that slice
produced by abstract  assignments may be larger than  the one produced
by  abstract  conditionals.  Given  a  program  $P$,  we shall  call  a
subprogram of  $P$ in which  some of the assignments  and conditionals
have been replaced by their  abstract versions as an {\em abstract 
subprogram}.

Now we will formally define a data slice and identify the statements which
should be part of the data slice.

\subsection{Data slice: Formalization}

Given  a slicing criterion  $\langle l,  V \rangle$,  a data  slice is
required to retain  the program's behavior in computing  the value set
of the  variables in  $V$, but is  not required  to visit $l$  as many
times  as  the original  program.  However,  whatever  value sets  are
computed  by the  slice should  match those  computed by  the original
program in a sense that we shall make precise now.

  Consider  program   P7   of
  Figure~\ref{fig:vic-example}.  Assume  that the number  of times the
  outer loop iterates depends on the input and the inner loop iterates
  a fixed  number of times, say  3.  If the outer  loop executes twice
  for  an  input  I, the  values  of  \texttt{x}  at  line 8  will  be
  3,6,9,3,6,9.  However, if both  the while conditions are replaced by
  `*' in a slice, then for the  same input, 3,3 is one of the sequence
  of values generated for \texttt{x}. This sequence does not match the
  sequence generated  by the concrete  program. On the other  hand, if
  only the  outer loop condition is  replaced by `*',  then the output
  produced will be  zero or more occurrences of  sequence 3,6,9. These
  sequences  are considered  to  match the  sequence  produced by  the
  original program.

Based  on  these considerations,   we identify   the
necessary properties of a data slice $P^D$ for a given program $P$ and slicing
criterion $\langle l, V \rangle$.
\begin{enumerate}
\item $P^D$ is an abstract subprogram of $P$
\item For every input I, on which augmented program terminates with trace $\tau$, there 
exists a trace  $\tau^{d}$ of $P^D$ on same input  I, such that $\tau$
and $\tau^{d}$ are {\em SC-equivalent}.
\item Let $\tau^{d}$ and $\tau$ be traces of $P^D$ and $P^A$
respectively   on  an   input   I.  Let   $k$  be the minimum of the numbers 
of {\em SC-execution} states in $\tau$ and
$\tau^{d}$.    Then  for  all  $i  \le  k$,  the  {\em V-restricted} program states of $i^{th}$
{\em SC-execution} states of $\tau$ and $\tau^{d}$ are the same.
\end{enumerate}
%% \sout{We elaborate third property formally as follows. Let $\tau^{d}$ be a trace
%% of program $P^D$  on an input $I$. Let $\tau$ be  the trace of program
%% $P^{A}$  on  same input  $I$.   Let  $i$ and  $j$  be  number of  {\em
%% SC-execution}  states  in  $\tau$  and $\tau^{d}$  respectively.   Let
%% $k_{th}( 0 <  k \leq min(i,j)$ {\em SC-execution}  state in $\tau$ and
%% $\tau^d$    be   $\langle    \Upsilon^l,    \sigma_k   \rangle$    and
%% $\langle  \Upsilon^l,  \sigma^d_k   \rangle$  respectively.  Then  the
%% property says that $\forall 0 < k \leq min(i,j) :
%% \lfloor \sigma_k \rfloor_{\Upsilon^V} = 
%% \lfloor \sigma^d_k \rfloor_{\Upsilon^V}$.} 
Clearly, a backward slice $P^B$ also satisfies the properties of 
data slice mentioned above. Therefore, $P^D \subseteq P^B$

Given  a slicing  criterion $\langle  l, V  \rangle$, we  now identify
statements  which are  necessarily in  the data  slice.  We  call such
statements {\em value-impacting} and define the term
shortly. Informally speaking,   a chain
of assignments that determine the value set of $V$ is value-impacting.
Further,  a condition  that  determines which  of  the several  values
generated  by  value-impacting  statements  reaches $l$  during
execution is also value-impacting. In subsequent 
discussions, we shall
 often use ``value-impacting statements'' to mean both 
value-impacting assignments and conditionals.

\begin{definition}{\em (Value-impacting statement)}
\label{vic_def}
A statement $s$ {\em value-impacts} $\Upsilon$, 
if any of the following conditions hold:
\begin{enumerate}
\item $s$ is the augmented $SKIP$ statement.
\item $s$ is an assignment,       and $s \in DU(\Upsilon)$.
\item $s$ is an assignment, and there exists  a statement $t$ such that
$t$ {\em value-impacts} $\Upsilon$ and $s \in DU(LV(t))$.
\item $s$ is a condition $c$,  and the following holds:
From $c$ there exist  paths $\pi_1, \pi_2$ to  $l$  starting from edges
$e_1$ and $e_2$ respectively. Further, there exists a statement $t$
such that $t$ {\em value-impacts} $\Upsilon$ and 
\begin{enumerate}
\item $t$ is the first value-impacting statement along $\pi_1$
\item $t$ is not the first value-impacting statement along  $\pi_2$.
\end{enumerate}
\end{enumerate}
\end{definition}

%{\color{magenta}

The triplet $\langle \pi_1, \pi_2, t\rangle$ due to which a 
condition $c$ satisfies rule (4) will be called 
a {\em witness} for a value-impacting condition.  Obviously, there can be 
more than one witnesses for a condition to be value-impacting. The set of all 
such witnesses will be referred as $WVI(c, \Upsilon)$.
%}

\begin{figure}
\centerline{\includegraphics[height=50mm, width=60mm]{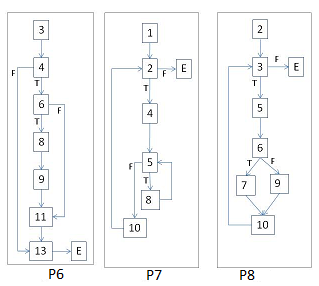}}
\caption{Control flow graphs}
\label{fig:vic-cfg}
\end{figure}

In   Figure~\ref{fig:vic-example}    we   show   some    examples   of
value-impacting  statements. The CFGs of these programs are shown in Figure~\ref{fig:vic-cfg}. In  P6,  lines  1  and  8  are
value-impacting for $\langle 11, \lbrace\mbox{\texttt{x}}\rbrace  \rangle$ because the values of \texttt{x}
generated at these statements reach  11.  In addition, condition \texttt{c2} is
also value-impacting for  the reason that of the two  paths from \texttt{c2} to
11, only one has \mbox{\texttt{x}}~=~\mbox{\texttt{z}}~+~5 as the first value-impacting statement. As a
consequence \texttt{c2}  determines whether  the value generated  at line 1  or at
line  8   reaches  11.    However,  notice  that   while  line   8  is
value-impacting  for  $\langle  9,  \lbrace\mbox{\texttt{x}}\rbrace\rangle$, condition  \texttt{c2}  is  not,
because  there is  no  path  to line  9  from the  false  edge of  \texttt{c2}.
Similarly, condition \texttt{c1} value-impacts $\langle 13,\lbrace\mbox{\texttt{x}}\rbrace\rangle$.

In P7 of  Figure~\ref{fig:vic-example}, the definition of \mbox{\texttt{x}} at lines 4 
and 7 are value-impacting
for $\langle 7,  \lbrace\mbox{\texttt{x}}\rbrace \rangle$.  Since line 7 is  not reachable along the
false  edge of  \texttt{c2} without  passing through  line 4,  \texttt{c2}  also becomes
value-impacting for $\langle 7, \lbrace\mbox{\texttt{x}}\rbrace  \rangle$.  In P8, we can see
that, \texttt{c2} is value-impacting for $\langle 7, \lbrace\mbox{\texttt{x}}\rbrace \rangle$.

Obviously, if a statement $s$ value-impacts a slicing criterion
$\Upsilon = \langle l, V \rangle$,  then $s$ can be
the cause of an erroneous value of some $v \in V$ at $l$
and should be examined while debugging. Therefore, $s$ must be part of
$P^D$.  
%\sout{Similarly,  if a statement $s'$ is  not value-impacting for
%$\Upsilon$ then $s{'}$ need not be in $P^D$}.

Let  $VI(\Upsilon)$  be  the  set  of  value-impacting  statements  of
$\Upsilon$.  Let $AC(\Upsilon)$ be conditional statements that are not
by  themselves  {\em   value-impacting},  but on which other value
impacting statements are strongly  (and transitively)
control dependent.  Formally:
$$AC(\Upsilon)  = (\bigcup_{t\in VI(\Upsilon)}\{c \mid c \longrightarrow t\})  \setminus  VI(\Upsilon)$$ 
We construct an abstract subprogram $P^S=VI(\Upsilon) \cup AC(\Upsilon)$
in which the  conditionals in $AC(\Upsilon)$ appear in an
abstract form.  Obviously, $P^S$  retains the structure of $P^A$ with
respect to all statements included in $P^S$.  We claim that $P^S$ is a
{\em data slice}.   To show this, we shall first  prove that the value
sets of $\Upsilon$ produced by execution of $P^S$ match those produced
by $P^A$.
\begin{lemma}
\label{vi_sat_dsp2}
Let $\tau$  and $\tau'$ be  traces of programs  $P^A$ and $P^S$  for an
input $I$.  Also assume that both  the traces go through  $l$ at least
once. Then the corresponding SC-execution states of $\tau$ and $\tau'$
will be the same when restricted to the variables in $V$.
\end{lemma}
\begin{IEEEproof}
Let   $\tau_{s}=[\langle  n_{i},   \sigma_{i}  \rangle],\; i \geq 0$  and
$\tau'_{s}  =  [\langle  n'_{j},\sigma'_{j}  \rangle],\; j \geq 0$  be  the
sequence of execution  states in $\tau$ and $\tau'$  
such that $n_0 = n'_0 = ENTRY$ and for $i>0$ and $j>0$ $n_{i},
n'_{j} \in VI(\Upsilon)$. Let $K$ be minimum of the number of elements
in  $\tau_{s}$ and  $\tau'_{s}$.  Since  $l$ occurs  at least  once in
$\tau$ and in $\tau'$, $K >  0$.  We will prove by induction on $i$
that for all $i \leq K$,  $n_{i} = n'_{i}$ and $\lfloor \sigma_{i}\rfloor_{Z} = 
\lfloor \sigma'_{i}\rfloor_{Z}$, where $Z = REF(n_{i})$.\\

{\em Base step  : i=0.} It holds trivially as $n_0=n'_0=ENTRY \wedge \sigma_0 = \sigma'_0 = I$.

{\em Induction  step:} Let the hypothesis  be true for $i  \leq K$ and
assume  that $i+1  \leq K$  (else the  proof holds  vacuously).  Since
$\lfloor\sigma_{i}\rfloor_{Z}  =  \lfloor  \sigma'_{i}\rfloor_{Z}$,
the  edge  followed  from   $n_{i}$  and  $n'_{i}$  in  $\tau$  and
$\tau'$ have  to be same.  
Assume that $n_{i+1}  \neq n'_{i+1}$.  
Let $c$ be  a common condition, with edges $e_1$ and $e_2$,
occurring between $n_{i}$ and   $n_{i+1}$  in $\tau$ and
between  $n'_{i}$ and $n'_{i+1}$  in $\tau'$.  
Clearly there  is such  a $c$,  otherwise $n_{i+1}$
would have been the same as  $n'_{i+1}$.  Since both
traces have occurrence  of $l$ and $n_{i+1}$ and  $n'_{i+1}$ are the first
value-impacting  statements  on  paths   from  $e_1$  and  $e_2$,  the
condition $c \in VI(\Upsilon)$ according to the definition.  This is 
contrary to  our assumption  that $n_{i+1}$ and  $n'_{i+1}$ are  the first
value-impacting statements in $\tau$  and $\tau'$ after $n_{i}$ and 
$n'_{i}$ respectively. Therefore, $n_{i+1} = n'_{i+1}$.

Now   suppose    that   for   some    variable   $x\in   Z$,
$\sigma_{i+1}(x)  \neq \sigma'_{i+1}(x)$.  Let $d$  be  statement which
provides value of  $x$ at $n_{i+1}$.  But then  $d \in VI(\Upsilon)$. If $d$ occurs before or at $n_i$ then it must be there in $\tau'$ also and 
therefore, $\sigma_{i+1}(x)  = \sigma'_{i+1}(x)$.  If $d$ occurs after 
$n_i$ then $\langle n_{i+1}, \sigma_{i+1}  \rangle$ can  not be  the first
element of $\tau_{s}$ after $\langle n_i, \sigma_i \rangle$. This is 
contrary to our assumption and therefore $\sigma_{i+1}(x)  = \sigma'_{i+1}(x)$.
\end{IEEEproof}

Now we  prove our claim  that $P^S$ is  a {\em data slice}.   We shall
show this by constructing a trace $\tau'$ for $P^S$ from a given trace
$\tau$  of  $P^A$ on  an  input $I$.   This  will  establish that  $P^S$
satisfies property (2) of data slice.  Using lemma~\ref{vi_sat_dsp2},
we shall show that $P^S$ satisfies property (3) as well.
\begin{theorem}
\label{vi_ss_ds}
The abstract  subprogram $P^S$  satisfies the property for  {\em data
slice}.
\end{theorem}
\newcommand{\colortext}[2][red]{{\color{#1} #2}}
\begin{IEEEproof}
Let $\tau$  be a trace for program  $P^A$ on input  $I$ that
has $K\geq      0$     execution     states.       Let     $\tau'=[\langle
n_{i}, \sigma_{i}  \rangle],\; 0 \leq i \leq K$, be  the sub-sequence of
$\tau$ such  that $n_{i}$,  $i \geq 0$
are nodes in CFG of $P^S$.  We show by induction on $i$ that for each $i \leq
K$, $[\langle n_{i}, \sigma_{i} \rangle]$  is also the prefix of a trace
for $P^S$.

{\em  Base step:  $i=0$.}  The lemma  holds
trivially as $n_{0} = ENTRY$.

{\em Induction step:} Assume that the hypothesis holds for some $i\leq
K$.  Let $n_{i}$ be a  condition.  If $n_{i} \in AC(\Upsilon)$ then it
is abstract  and can take  either branch. If $n_{i}  \in VI(\Upsilon)$
then by lemma~\ref{vi_sat_dsp2} it will have the same  value as in trace
$\tau$.  So  for any  edge taken out  of $n_i$  in $\tau$, there  is a
trace of $P^S$ which takes the same edge.

Now assume  that for none of  the traces of $P^S$  is the $(i+1)^{th}$
node same as  $n_{i+1}$. This must be because of some condition
$c$ before $n_{i+1}$, but after $n_i$, in $\tau$.    
But  then  $c  \longrightarrow n_{i+1}$ and therefore $c \in P^S$. So
$\langle n_{i+1}, \sigma_{i+1} \rangle$ can not be the first execution state
after $\langle n_{i}, \sigma_{i} \rangle$ in $\tau'$, a contradiction. 

Thus, property  (3) is  satisfied.  By lemma~\ref{vi_sat_dsp2},  it is
obvious that property (2) is also satisfied.
\end{IEEEproof}
%}
%{\color{blue}
\subsection{Computing data slice using data and control flow}

%\begin{figure}[!htb]
\begin{figure}
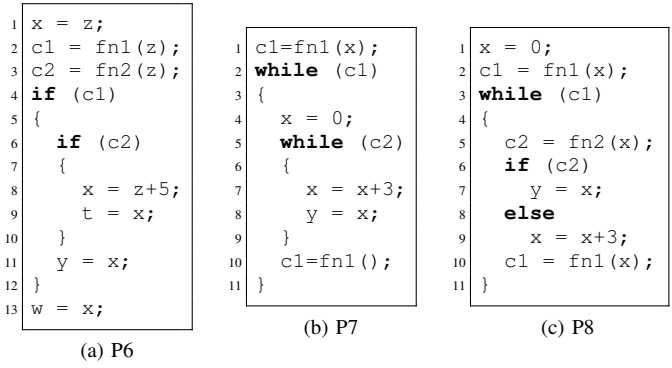

\begin{minipage}{0.80in}
\subfloat[P6]{%
\begin{lstlisting}[style=MyListStyle]
x = z;
c1 = fn1(z);
c2 = fn2(z);
if (c1)
{
  if (c2)
  {
    x = z+5;
    t = x;
  }
  y = x;
}
w = x;
\end{lstlisting}%
}
\end{minipage}
\hspace{0.7cm}
\begin{minipage}{0.80in}
\subfloat[P7]{%
\begin{lstlisting}[style=MyListStyle]
c1=fn1(x);
while (c1)
{
  x = 0;
  while (c2)
  {
    x = x+3;
    y = x;
  }
  c1=fn1();
}
\end{lstlisting}%
}
\end{minipage}
\hspace{0.7cm}
\begin{minipage}{0.95in}
\subfloat[P8]{%
\begin{lstlisting}[style=MyListStyle]
x = 0;
c1 = fn1(x);
while (c1)
{
  c2 = fn2(x);
  if (c2)
    y = x;
  else
    x = x+3;
  c1 = fn1(x);
}
\end{lstlisting}%
}
\end{minipage}
\caption{Programs to explain value-impact}
\label{fig:vic-example}
\end{figure}

\begin{figure}
\centerline{\includegraphics[height=80mm, width=80mm]{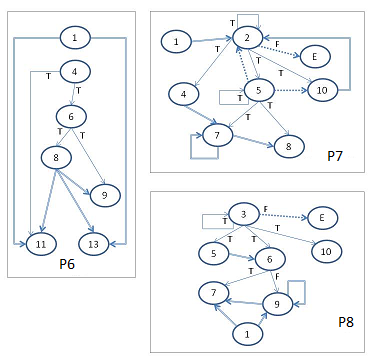}}
\caption{Program dependence graphs}
\label{fig:vic-pdg}
\end{figure}

We  now relate  $VI(\Upsilon)$  to control  and  data dependence.   In 
Figure~\ref{fig:vic-pdg}, we show PDG for programs of 
Fig.~\ref{fig:vic-example}. Solid, normal and dotted arrows show data 
dependence, strong control dependence and weak 
control dependence respectively.
In P6  of Fig.~\ref{fig:vic-example}, \texttt{c2}  is value-impacting for
$\langle 11, \lbrace\mbox{\texttt{x}}\rbrace  \rangle$.  In terms of control-dependence,  line 11 is
not control dependent on  \texttt{c2} while the value-impacting assignment at
line   8  is.   For   exactly  similar   reasons,   condition  \texttt{c1}   is
value-impacting  for  $\langle   13,  \lbrace\mbox{\texttt{x}}\rbrace\rangle$.   We  generalize  the
identified condition  for a slicing  criterion $\Upsilon =  \langle l,
V  \rangle$  as  {\em  $cond_1$}:  $l$ is  not  control  dependent  on
condition  $c$  and  a  value-impacting statement  for  $\Upsilon$  is
transitively control dependent on $c$.

In  P7,  \texttt{c2} is  value-impacting  for $\langle  7, \lbrace\mbox{\texttt{x}}\rbrace\rangle$,  a
slicing   criterion  whose   program  point   (line  7)   is  strongly
true-control dependent on \texttt{c2}.  The value-impacting assignment for this
criterion  at  line  4   is  weakly  false-transitively control  dependent  on  \texttt{c2}.
Similarly, in  the case of  P8, condition \texttt{c2}  is value-impacting
for $\langle 7, \lbrace\mbox{\texttt{x}}\rbrace\rangle$ and  7 is strongly true-control dependent on
\texttt{c2}.  Moreover,  the value-impacting assignment  at line 9  is strongly
false-control dependent on \texttt{c2}.  From these observations, we identify a
second condition $cond_2$: $l$ is  control dependent on $c$ and $l$ is
reachable from  one edge and a value-impacting  statement is reachable
from other edge. We show that the disjunction of $cond_1$ and $cond_2$
is  a  necessary condition  for  value-impact.  Thus  we can  use  the
disjunction  to  compute  an  over-approximation  of  value  impacting
conditions.  
%}

As  seen from  the examples,  to  capture value-impact  we need  to
consider both strong and weak control dependence. For this we make use of
transitive control dependence.
%\begin{definition}{(Transitive control dependence)}
%For given statement $s$ and condition $c$, if there is a path $\pi$ in
%the  program  dependence  graph  (PDG)  [reference] from  $c$  to  $s$
%consisting of  weak or  strong control dependence  edges, then  $s$ is
%said to be {\em transitively  control dependent} on $c$ and denoted as
%$c  \rightsquigarrow s$.  However,  if $\pi$  consists of  only strong
%control dependence  edges, then  $s$ is said  to be  {\em transitively
%strong control dependent} on $c$ and denoted as $c \longrightarrow s$.
%Further, if $\pi$  above starts with a control  dependent edge labeled
%$e$,   we   may  highlight   this   by   writing   $c  \stackrel   {e}
%{\rightsquigarrow} s$ ($c \stackrel {e} {\longrightarrow} s$).

We now make the following connections between 
transitive control dependence and weak post dominance.
\begin{claim}
\label{scd_semctrl}
Given a  condition $c$ and a  statement $s$, assume  that $c \stackrel
{e} \longrightarrow s$. If there  is a statement $t$ distinct from $c$
such  that there is  a path  from $c$  to $s$  starting from  the edge
$\bar{e} \neq  e$ and going  through $t$, then $c  \stackrel {\bar{e}}
{\rightsquigarrow} t$.
\end{claim}
\begin{claim}
\label{postdom_semctrl}
Let $c$ be  a condition and $u$ be a statement  distinct from $c$ such
that  $u$ is  the immediate  post-dominator of  $c$.  If  there is  a
statement $t$ distinct from $c$  and $u$ such that  there is a path
from $c$ to  $u$ starting with edge $e$ and  passing through $t$, then
$c \stackrel {e} {\longrightarrow} t$.
\end{claim}

We  shall  now  establish  that  if  a condition  $c$  is  {\em  value-impacting}, then there must  be another {\em value-impacting} statement
which is {\em transitively-control dependent} on $c$.

\begin{lemma}
\label{vi_semctrl}
Assume that for a slicing criterion $\Upsilon = \langle l, V \rangle$,
 $c \in VI(\Upsilon)$. Then one of the following holds:

\noindent $\exists 
\langle \pi_1, \pi_2, t\rangle \in WVI(c, \Upsilon):
    (\neg (c \longrightarrow l) \wedge c \longrightarrow
    t),\,\, \mbox{or}$\\
$\exists 
\langle \pi_1, \pi_2, t\rangle \in WVI(c, \Upsilon):
(c \stackrel {e_2} {\longrightarrow} l \wedge
\neg(c \stackrel {e_2} {\longrightarrow} t) \wedge
     c \stackrel {e_1} {\rightsquigarrow} t)$.

\noindent where $e_1$ and $e_2$ are starting edges of paths $\pi_1$ and $\pi_2$ 
respectively.
\end{lemma}
\begin{IEEEproof}
%% Let $\langle \bar{\pi_1}, \bar{\pi_2}, \bar{t}\rangle \in WVI(c, \Upsilon)$.
%% Obviously $\bar{t}\neq c$. Let $\bar{e_1}$ and $\bar{e_2}$ be the starting
%% edges of $\bar{\pi_1}$ and $\bar{\pi_2}$ respectively. We consider
%% following cases:
Let $c \in VI(\Upsilon)$. The proof is by case analysis:

{\em Case 1:} Assume $\neg (c \longrightarrow l)$. Then there
must exist  $u \neq c$  such that $u$  is immediate post  dominator of
$c$. Since  $c \in VI(\Upsilon)$, by definition~\ref{vic_def} there are
paths  $\pi_1$  and  $\pi_2$  starting  from $e_1$  and  $e_2$  and  a
value-impacting statement $t$ such that  at least one of the ${\pi_1}$
and ${\pi_2}$ should have $t$  before $u$. Without loss of generality,
assume  that  $t$  is on $\pi_1$. Then  by  claim~\ref{postdom_semctrl},
$c \longrightarrow t$. Thus we  have proved that for the witness
$\langle  {\pi_1}, {\pi_2},  {t}\rangle \in  WVI(c,  \Upsilon)$, $\neg
(c \longrightarrow l)$ and $c \longrightarrow {t}$.

{\em Case  2:} Assume $c  {\longrightarrow} l$, and call  the starting
edge in  the chain of  control dependence from  $c$ to $l$  as $e_2$,
i.e.  $c \stackrel {e_2}  {\longrightarrow} l$. From the definition of
value-impacting condition, there is a  path from $c$ to $l$ that has a
statement, say $t$, as the first value-impacting statement and another
that  does not have  $t$ as  the first  value-impacting  statement. Now
there are two cases:

1. Let the path that has $t$ as the first value-impacting statement
leave  $c$ through  the edge  $e_1$. For  the kind  of  programs under
consideration, no  path from $c$  to $l$ can bypass  $e_2$.  Therefore
${t}$ on ${\pi_1}$  must be between ${e_1}$ and  ${e_2}$. Thus we have
$~\neg   (c  \stackrel   {{e_2}}  {\longrightarrow}   {t})$,   and  by
claim~\ref{scd_semctrl}, $c \stackrel {{e_1}} {\rightsquigarrow}
{t}$. The witness in this case being $\langle   {\pi_1},  {\pi_2},
{t}\rangle$. 

2. Now let the path that has $t$ as the first value-impacting statement
leave $c$ through  the edge $e_2$. Once again the  path that goes from
$c$ to  $l$ through $e_1$  must go through  $e_2$ and there must  be a
first value-impacting  statement $w$ between  $e_1$ and $e_2$.  Now we
have a witness $\langle {\pi_1}, {\pi_2}, {w}\rangle$ for which $~\neg
(c    \stackrel    {{e_2}}    {\longrightarrow}    {w})$,    and    by
claim~\ref{scd_semctrl}, $c \stackrel {{e_1}} {\rightsquigarrow} {w}$.
\end{IEEEproof}
Based  on  lemma~\ref{vi_semctrl},  we  shall  give  criteria  for
computing a  set of  statements \CVI, ({\em  computed value-impacting}
statements).  \CVIonly\    uses data  and  control dependence  and
computes an over-approximation of $\mathit{VI}$.

A statement $s$ is in \CVI\ if:
\begin{enumerate}
\item $s$ is the augmented $SKIP$ statement at $l$.
\item $s$ is an assignment such that $s \in DU(\Upsilon)$.
\item $s$ is an assignment,  and there is a statement $t$ such that
$t \in \CVI$ and $s \in DU(LV(t))$.
\item $s$ is a condition $c$, with two outgoing edges, labeled as $e_1$ and $e_2$, and $\exists
t \in \CVI$ satisfying one of the following:
\begin{enumerate}
\item $c \longrightarrow t \wedge \neg(c \longrightarrow l)$
\item $c \stackrel{e_1}{\longrightarrow} l  \wedge \neg( c \stackrel {e_1} {\longrightarrow} t ) \wedge c \stackrel{e_2}{\rightsquigarrow} t$
\end{enumerate}
\end{enumerate}
Condition 4) of \CVIonly\  is  motivated by lemma~\ref{vi_semctrl}.

We shall  now show that \CVI\   computes an over  approximation of value-impacting  statements  $VI(\Upsilon)$. In  subsequent  proofs, we  use
$MLP(s_1, s_2)$ to denote maximum length  of a loop free path in a CFG
from $s_1$ to $s_2$. In particular, $MLP(s,s)=0$.

\begin{lemma}
\label{vi_ss_tif}
$VI(\Upsilon) \subseteq \CVI$
\end{lemma}
\begin{IEEEproof}
Let $s  \in VI(\Upsilon)$.  We will  prove the result  by induction on
$MLP(s, l)$.

{\em  Base step:}  $MLP(s, l)=0$.   $s$ must  be the augmented $SKIP$
statement and   therefore in \CVI.

{\em   Induction    step:}   Let   the   hypothesis    be   true   for
all $s$ such that $MLP(s, l)\leq i$ and consider a $s$
for which $MLP(s, l) = i+1$ and $s \in
VI(\Upsilon)$.  If $s$ is an assignment, then  $s  \in \CVI$ from
definition. Let $s$  be a condition  $c$.  Let $\langle  \pi_1, \pi_2,
t\rangle   \in    WVI(c,   \Upsilon)$   satisfy    the   criteria   of
lemma~\ref{vi_semctrl}.   Obviously  $MLP(t, l)   \leq  i$.
and  $t \in \CVI$ by the induction hypothesis.  By definition, $c \in
\CVI$.
\end{IEEEproof}

We will now show that though \CVIonly\  is an over-approximation
of $\mathit{VI}$, it is contained  within the backward slice $P^B$. 
\begin{lemma}
\label{dfi_ss_bs}
$\CVI \subset  P^B_{\Upsilon}$.
\end{lemma}
\begin{IEEEproof}
Once again the proof is by induction on $MLP(s,l)$.

{\em Base case:} $i=0$. $s$ must be the augmented $SKIP$ statement and
 and therefore $s \in P^B$

{\em Induction step : }
Let   the   hypothesis    be   true   for
all $s$ such that $MLP(s, l)\leq i$ and consider a $s$
for which $MLP(s, l) = i+1$ and $s \in
\CVI$. By definition of \CVI, we have the following cases:
\begin{enumerate}
\item $s$ is an assignment and $s \in DU(\Upsilon)$. Clearly $s \in P^B$.
\item $s$ is assignment and $\exists t\in \CVI:\, s \in DU(LV(t))$. 
Clearly  $MLP(t,l) \leq  i$  and therefore  $t  \in P^B$.  By
construction of backward slice,  $s \in P^B$.
\item $s$ is a condition $c$, say with two edges $e_1$ and $e_2$. 
By definition of \CVI, $\exists t: t \in \CVI$, satisfying one of the following:

{\em Case a:} $c \longrightarrow t \wedge \neg(c \longrightarrow l)$.
Obviously,  $MLP(t, l)  \leq i$  and  $c \longrightarrow  t$.  By  the
induction hypothesis, $t  \in P^B$.  Therefore  $c \in
P^B$.

{\em  Case  b:}  $c  \stackrel{e_1}{\longrightarrow}  l  \wedge  \neg(
c     \stackrel     {e_1}     {\longrightarrow}     t     )     \wedge
c \stackrel{e_2}{\rightsquigarrow} t$.
In this case $c \longrightarrow l$ and therefore, by construction of
backward slice,  $c \in P^B$.
\end{enumerate}
\end{IEEEproof}

\section{Data slice computation}

As stated earlier,  \CVI\ provides the core set  of statements of the  data
slice. To make the slice executable, conditions are added using one of
the  methods of creating  abstract sub  programs described  before. We
present  an  algorithm  to  compute $\mathit{CVI}$  only;  adding  the
abstract  conditions  and   assignments  is  straightforward.   As  an
example, the conditions $c$ to  be abstracted are given by $\lbrace c\mid
c \longrightarrow l\rbrace \backslash \CVI$.

%\textbf{DDS} consists of statements $DFI(\Upsilon)$.
%
%\textbf{CDS} consists of conditionals,which were not included in DDS, 
%    and on which $l$ is {\em transitively strongly control dependent}.
%
%\textbf{ADS} consists of abstract
%form of those assignments, which are not part of DDS, and which directly 
%provide value to conditionals of CDS. 
%
%\textbf{ACS} consists of abstract conditionals, which were not included in 
%DDS or CDS, and on which, assignments included in ADS are {\em transitively 
%strongly control dependent}.
%
%We define computation of $CDS, ADS$ and $ACS$ when we use assignment
%abstraction as follows:\\
%\\
%%\begin{description}
%$CDS = \mbox{ {\em sconds*}}(l) - DDS$ \\
%$ASS = \bigcup_{s \in CDS} DU(LV(s)) - DDS$ \\
%$ADS = abstract(ASS)$ \\
%$DC = (\bigcup_{s \in ASS}\mbox{ {\em sconds*}}(s)) - (DDS \cup CDS) $ \\
%$ACS = abstract(DC)$ \\
%\end{description}

%To compute these when we use condition abstraction, we directly put CDS 
%in abstract form and ASS,ADS and ACS are not computed. Computation of $CDS, ADS$ and $ACS$ is straight forward and so we will not present detailed
%algorithm for the same.
%
%Finally the data slice, $P^D$, is as follows:
%
%$P^D = DDS \cup CDS \cup ADS \cup ACS$

\begin{algorithm}
\caption{Identifying $\mathit{CVI}$  conditions}
\label{alg:infconds}
\begin{algorithmic}[1]
\STATE \textbf{procedure} getCVIConds(t, lconds)
\STATE begin
\STATE $R = \emptyset$
\STATE $tconds = tcntrls(t)$
\FORALL {conditions $c$ appearing in $tconds$}
    \STATE $lsttab[T] = (\langle c,T,true\rangle \in lconds)$
    \STATE $lsttab[F] = (\langle c,F,true\rangle \in lconds)$
    \STATE $tsttab[T] = (\langle c,T,true\rangle \in tconds)$
    \STATE $tsttab[F] = (\langle c,F,true\rangle \in tconds)$
    \STATE $tswtab[T] = (\langle c,T,false\rangle \in tconds)$
    \STATE $tswtab[F] = (\langle c,F,false\rangle \in tconds)$
    \IF {$(\neg lsttab[T] \wedge \neg lsttab[F] \wedge ( tsttab[T] \vee tsttab[F]))$}
        \STATE add c to R
    \ELSE
        \IF {$((lsttab[T] \wedge \neg tsttab[T] \wedge tswtab[F]))
            \vee (lsttab[F] \wedge \neg tsttab[F] \wedge tswtab[T])$}
                \STATE add c to R
        \ENDIF
    \ENDIF
\ENDFOR
\STATE return R
\STATE end

\end{algorithmic}
\end{algorithm}

\subsection{Computing \CVI}
In computing \CVI,  the critical part is to  identify conditional expressions which
satisfy criteria for being in  \CVIonly.
%$ for a given statement $t$ which is
%already in $DFI(\Upsilon)$.  
We assume that  the $PDG$ already exists with  weak and strong control
dependences  and  data dependences.   Such  a  $PDG$  can be  computed
efficiently by algorithm  of Bilardi and Pingali ~\cite{COMP_WEAK_CD}.
From  the PDG,  for  a given   statement  $s$,  we can  find the  set
$conds(s)$ of pairs  $\langle c, e, b\rangle$, 
such  that $c \stackrel{e} {\rightsquigarrow} s$ when  $b=false$ 
and $c \stackrel{e} {\longrightarrow} s$ when  $b=true$.   Using
$conds(s)$,  we can compute the  set of  conditions on  which $s$  is {\em
  transitively control depdendent}.  We  call this set as $\mbox{ {\em
    tcntrls}}(s)$.   We do  so  by
traversing the PDG  and taking a transitive closure  of $conds(s)$. By
the definition  of \CVI,  we need to  examine only the  conditions $c$
which  appear in $\mbox{ {\em tcntrls}}(t)$ for  a given  statement $t$.
Algorithm  ~\ref{alg:infconds} computes  the set  of  conditions which
satisfy  the criteria for  \CVI\ for  a given  statement $t$  which is
already in \CVI. 
Lines 6 to  11 identify the kinds of  transitive  control dependence that $t$
has  on the outgoing edges of $c$.  
While  {\em tsttab[e]  = true}  means $c  \stackrel {e}
{\longrightarrow} t$,  {\em tswtab[e] =  true} means $c  \stackrel {e}
{\rightsquigarrow}  t$.   Algorithm  ~\ref{alg:cdsComp}  computes  the
complete set \CVI\  using a worklist based approach.  A node comes on
the  worklist only once.   The final  result is  denoted by  a boolean
array $inslice$  having value $true$  for every statement  included in
\CVI.

\begin{algorithm}
\caption{Computation of \CVI}
\label{alg:cdsComp}
\begin{algorithmic}[1]
\STATE \textbf{procedure} computeCVISet(l, V)
\STATE begin
\STATE initialize $inslice$, $inwl$ with $false$
\STATE $inslice[l] = true$ ; $lconds = tcntrls(l)$
\STATE $wl = \{\}$
\STATE $duset = DU(l, V)$
\FORALL { $s \in duset$ }
    \STATE add s to wl ; $inwl[s] = true$
\ENDFOR
\WHILE{ wl is not empty}
    \STATE remove next element w from wl
    \STATE $inslice[w] = true$
    \STATE $duset = DU(LV(w))$ 
    \STATE $cset = getCVICond(w, lconds)$
    \FORALL { $ s \in (duset \cup cset)$ }
        \IF {$inwl[s] = false$}
        \STATE add s to wl ; $inwl[s] = true$
        \ENDIF
    \ENDFOR
\ENDWHILE
\STATE end

\end{algorithmic}
\end{algorithm}

%\begin{algorithm}
%\caption{Addition of abstract conditions}
%\label{alg:absConds}
%\begin{algorithmic}[1]
%\input{alg_absConds}
%\end{algorithmic}
%\end{algorithm}

%\begin{algorithm}
%\caption{Addition of abstract statements}
%\label{alg:absStmts}
%\begin{algorithmic}[1]
%\input{alg_absStmts}
%\end{algorithmic}
%\end{algorithm}
\subsection{Algorithm complexity}
Assume  there are  $N$ nodes,  $E^d$  data dependent  edges and  $E^c$
control dependent  edges, giving a  total of $E=E^d+E^c$ edges  in the
PDG.  In  $getCVIConds$, computing $tcntrls$ for the  given node, will
take $O(E^c)$ time.  The checks in lines 6 to 11 can be done in $O(1)$
time with a  space complexity of $O(N)$.  Since the  checks have to be
made  for  all  conditions  occurring  in $tcntrls$,  the  worst  case
complexity  of $getCVIConds$  would be  $O(N +  E^c)$.   In algorithm,
$computeCVISet$,  a node  goes in  the worklist  only  once, therefore
there would  be maximum $N$ invocations of $getCVIConds$.   The worst case
complexity  of entire  algorithm  is  $O(E^c \times  N  + N^2  +E^d)$.
However  in practice,  the loop  at line  5 in  $getCVIConds$  will be
executed much  fewer times than $N$  and nodes going  in worklist will
also be of  the size of the data slice.  As  a result, the algorithm's
average  complexity  will  be  $O(N  + E^c+E^d)$.   In  contrast,  the
backward slice is  computed in $O(E)$ time, in  worst case.  Note that
in both cases the time complexity is arrived at by assuming that $PDG$
have already  been built.   In practice, our  results have  also shown
that there is only a marginal increase in time in computing data slice
from  that taken  in backward  slice when  compared to  time  taken in
building the PDG itself.
 %4
\section{Implementation and measurements}

We implemented the algorithm to compute data slices using our in-house 
data flow analysis framework called PRISM which is based on
the JAVA platform. It can 
construct PDGs and can compute the conventional backward slices. 
It  has been used for developing static analysis tools~\cite{TECA_EXPERIENCE, FALSE_POSITIVE_FILTERING}.
We have used a context and flow sensitive points-to analysis. 
The backward slicing algorithm is also context sensitive and field sensitive. 
Thus it represents the state of the art in backward slicing.
We computed data slice using  condition abstraction approach 
which is suitable for debugging and program understanding. 

\begin{figure*}[t]
\centering
\begin{tabular}{|l|c|c|c|c|c|c|c|c|c|c|c|} \hline\hline
%\multicolumn{1}{|c}{} & \multicolumn{1}{|c}{} & \multicolumn{1}{|c}{} & \multicolumn{9}{|c|}{Average Sizes} \\ \cline{4-12}
%\multicolumn{1}{|c}{} & \multicolumn{1}{|c}{} & \multicolumn{1}{|c}{} & \multicolumn{3}{|c}{Average size(in nodes)}  & \multicolumn{3}{|c}{Average size (as \%of program)} & \multicolumn{3}{|c|}{Time (in seconds)}\\ 
\multicolumn{1}{|c}{} & \multicolumn{1}{|c}{Size} & \multicolumn{1}{|c}{number of} & \multicolumn{3}{|c}{Avg. Size (in nodes)}  & \multicolumn{3}{|c}{Avg Size as \% of program} &
\multicolumn{3}{|c|}{time in seconds}\\ \cline{4-12}
Program & (nodes) & slices & BS & DS & CS & BS & DS & CS & BS & DS & CS \\ \hline
(1) & (2) & (3) & (4) & (5) & (6) & (7) & (8) & (9) & (10) & (11) & (12) \\ \hline
1 & 1063 & 10 & 229 & 60 & 227&  21.63 &   5.65 &  21.37&  26.33 &  26.34 &  26.32  \\ \hline
2 & 1183 & 10 & 266 & 123 & 240&  22.49 &  10.43 &  20.36&  32.05 &  32.07 &  32.04  \\ \hline
3 & 2944 & 10 & 360 & 257 & 236&  12.26 &   8.75 &   8.03&  38.15 &  38.21 &  38.13  \\ \hline
4 & 881 & 10 & 148 & 62 & 148&  16.89 &   7.12 &  16.81&  60.43 &  60.47 &  60.42  \\ \hline
5 & 1607 & 10 & 226 & 137 & 203&  14.08 &   8.55 &  12.65&  39.46 &  39.51 &  39.45  \\ \hline
6 & 2246 & 10 & 447 & 348 & 444&  19.93 &  15.49 &  19.79&  56.47 &  57.24 &  56.45  \\ \hline
7 & 2493 & 5 & 167 & 81 & 163&   6.70 &   3.26 &   6.54&  51.45 &  51.48 &  51.42  \\ \hline
8 & 2635 & 10 & 842 & 257 & 842&  31.98 &   9.79 &  31.97&  44.20 &  44.31 &  44.19  \\ \hline
9 & 2992 & 9 & 437 & 149 & 429&  14.63 &   5.01 &  14.34&  52.67 &  52.76 &  52.65  \\ \hline
10 & 1625 & 10 & 190 & 94 & 178&  11.70 &   5.84 &  10.98&  63.80 &  63.83 &  63.79  \\ \hline
11 & 3413 & 10 & 733 & 341 & 728&  21.50 &  10.01 &  21.35&  92.74 &  92.97 &  92.70  \\ \hline
12 & 3105 & 5 & 571 & 412 & 563&  18.41 &  13.29 &  18.13&  71.41 &  89.70 &  71.35  \\ \hline
13 & 4452 & 1 & 369 & 70 & 317&   8.29 &   1.57 &   7.12& 102.29 & 102.27 & 102.26  \\ \hline
14 & 5236 & 1 & 982 & 25 & 982&  18.75 &   0.48 &  18.75& 116.92 & 116.47 & 116.53  \\ \hline
15 & 2616 & 10 & 948 & 761 & 945&  36.27 &  29.12 &  36.15& 208.23 & 230.40 & 208.24  \\ \hline
16 & 3883 & 10 & 1202 & 180 & 1202&  30.97 &   4.64 &  30.96& 215.99 & 216.14 & 215.97  \\ \hline
17 & 802 & 10 & 92 & 56 & 90&  11.48 &   7.01 &  11.28&  53.85 &  53.86 &  53.84  \\ \hline
18 & 8116 & 10 & 3489 & 1637 & 3143&  42.99 &  20.18 &  38.73& 447.69 & 512.58 & 447.30  \\ \hline
19 & 6746 & 10 & 1928 & 1558 & 1923&  28.58 &  23.10 &  28.51& 272.10 & 293.40 & 271.74  \\ \hline
20 & 11104 & 10 & 4096 & 1214 & 4053&  36.89 &  10.94 &  36.50& 301.26 & 322.31 & 301.13  \\ \hline

Overall & & & & & & 24.89 & 10.66 & 24.04 & & &\\ \hline
\end{tabular}

\caption{Average slice sizes and computation time}
\label{fig:experiment-tab1}
\end{figure*}

Although we have described our algorithm at an intra-procedural level,
our implementation performs interprocedural analysis by
summarising procedure calls by a sequence of assignments 
simulating the
use-def summary of called procedure. We trigger
additional slicing criteria at call points based upon the values needed
at procedure entry point in a context sensitive manner. We computed data 
slice for these additional slicing criteria and at the end took a union
of all of them. This may introduce some imprecision but is sound.

Our experiments have been carried out on 3.0 GHz Intel Core2Duo processor 
with 2 GB RAM and 32 bit OS.
Measurements  were peformed on 42 modules of varying sizes of a proprietary code base of a large
navigational system of an automobile.
Due to space constraints, we have presented summary data of only 20 modules
in Figure~\ref{fig:experiment-tab1}. 
Column (1) gives anonymized program names. 
Column (2) lists the number of CFG nodes. 
Column (3) gives number of slices computed for the program. For this purpose 
we randomly selected the return statements of functions returning some
value as the slicing criterion. For each program, we picked up maximum ten such slicing
criteria.
In all, we created slices for total 391 such slicing criteria spread over 42 programs.

Columns (4), (5) and (6) provide the average sizes of slices in terms of the number of 
nodes for backward slice (BS), data slice (DS) and control slice (CS) 
respectively. Columns (7), (8) and (9) provide the average sizes of slices as 
a percentage of the program size (as given in column (3)) for BS, DS and CS respectively.
The average time taken (in seconds) in computing these slices is shown
in columns (10), (11) and (12) in same order. This time includes the time taken 
for constructing PDGs.

\begin{figure}
\includegraphics[scale=0.54]{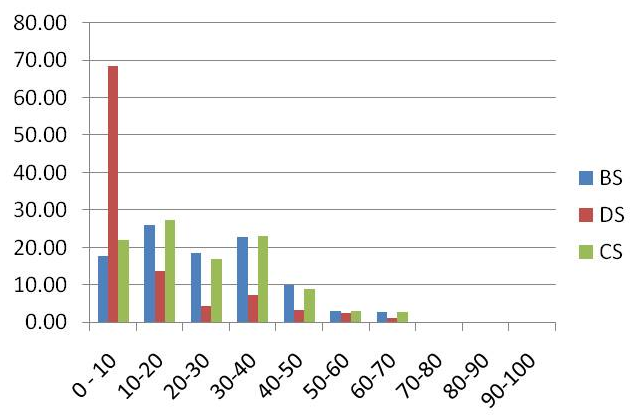}
\caption{Slice size distribution. {\em X-axis} shows slice size as percentage of program size and {\em Y-axis} shows percentage of slicing criteria for which this size was observed.}
\label{fig:ss_dist_graph}
\end{figure}

Figure~\ref{fig:ss_dist_graph} shows the distribution of sizes of slices in terms of percentage of program sizes through a graph. 
The X axis shows the
size of the slice and Y axis shows the percentage of slicing criteria exhibiting 
that size. In the figure, \emph{BS, DS} and \emph{CS} stand for {\em backward slice, data slice} and \emph{control slice} respectively. 
It is clear that in more than 60\% of the cases, the size of a data slice is smaller than 10\% of the code size.  Besides,
the time required to compute a data slice  is comparable to the time  taken for computing a backward slice.

Note that the average size of backward slices is 
25\% of the code size which matches the observation by Binkley et al.~\cite{SLICE_SIZES}.
suggesting our slices are comparable in precision. Average size 
of data slice is found to be 10\% of the code size. Given such a reduction, data slices
may be very helpful in
debugging,
property checking and program understanding. 
This data corroborates our intuition that most statement are included in a
backward slice because they influence the reachability of the slicing criterion
rather than the value computed. It is not surprising then that
the size of a control slice is comparable to that of the corresponding backward
slice in majority of the cases.
The average size of 
control slice turns out to be of 24.4\% of program size which is comparable to the size of backward 
slice (24.89\%).
 %5
%\input{relatedwork} %6
\section{Conclusion}

Different applications of program understanding require different
combinations of influences governing data computations and control flow.
For example, in the case of debugging for wrong output values, the influences governing the
reachability of the statement of interest are irrelevant. 

It follows that separating the influences of data and control in a 
backward slice by constructing separate data and control slices is an
effective way of producing smaller programs for debugging, 
program understanding and property checking. 
In the case of 
debugging for wrong output values, 
a data slice provides a much smaller piece of code to investigate than that provided by a backward slice for the same slicing criterion. 

We have provided formal definitions of data and control slices, defined algorithms to compute them, have shown the soundness of the
algorithms, and have presented the results of our empirical experiments.
Our measurements show that a data slice is indeed 
much smaller than the corresponding backward slice and is computable in comparable time. 

In future, we would like to investigate the minimality of data slices and efficient algorithms to compute them.
We would also like to explore the effectiveness of data slices for much larger programs.
%An empirical study of
%data slice computation and its usability on large programs can also be undertaken.

\balance

\bibliographystyle{plain}
\bibliography{slices}

\end{document}